\documentclass[acus]{JAC2000}

\usepackage{graphicx}

\begin{document}
\title{LUMINOSITY OF ASYMMETRIC e+e- COLLIDER WITH COUPLING LATTICES
    \thanks{Work supported by the Department of Energy under Contract No.
    DE-AC03-76SF00515}}

\author{ Yunhai Cai, SLAC, Stanford, CA 94309, USA }

\maketitle

\begin{abstract} 
    A formula of luminosity of asymmetric e+e- collider with coupled 
lattices is derived explicitly. The calculation shows how the tilted
angle and aspect ratio of the beams affect the luminosity. Knowing the
result of the calculation, we measure the tilted angle of the luminous 
region of the collision for PEP-II using two dimensional transverse 
scan of the luminosity. The method could be applied to correct the 
coupling at the collision point. 

\end{abstract}

\section{INTRODUCTION}
    The luminosity of colliding storage rings is one of the most 
important criteria of performance. In order to achieve high luminosity,
the colliding beams need to be aligned precisely at the collision point 
in all three dimensions. For a symmetric collider, the alignment of the 
electron and positron beams is ensured automatically by the symmetries of 
charge conjugation and time-reversal. Since an asymmetric collider 
consists of two different storage rings, precise alignment of the two 
beams at the collision can only be achieved through tuning of the two rings.

    In this paper, our goal is to show some effects on the luminosity
when the two beams are not aligned well in transverse planes due to the 
coupling.
   
\section{LINEAR COUPLING}
It was shown by Edwards and Teng\cite{edwards} that a two-dimensional 
coupled linear motion in a periodic and symplectic system can be 
parameterized with ten independent parameters as a block diagonalization
of a one-turn matrix
\begin{equation}
M = A \cdot R \cdot A^{-1} 
\end{equation}
where $R$ is the rotation matrix and $A$ defines the symplectic transformation
from the normalized coordinates to the physical coordinates. These matrices
can be further decomposed into
\begin{equation}
R =  \left( \begin{array}{clcr} r_1 & 0 \\ 
                               0 & r_2 
           \end{array} \right) 
\end{equation}
and
\begin{equation}
A = \left( \begin{array}{clcr} I \cos\phi & {\bar w} \sin\phi \\ 
                               -w \sin\phi & I \cos\phi 
           \end{array} \right)
    \left( \begin{array}{clcr} s_1 & 0 \\ 
                               0 & s_2 
           \end{array} \right)
\end{equation}
where $I$ is 2$\times$2 identity matrix and $s_1,s_2,w$ are 2$\times$2 
symplectic matrices with a determinant of unity. The angle $\phi$ is called 
coupling angle. Here we denote the bar as two-dimensional symplectic conjugate
\begin{equation}
{\bar w} = - J \cdot w^T \cdot J
\end{equation}
where $J$ is 2$\times$2 unit symplectic matrix
\begin{equation}
J = \left( \begin{array}{clcr} 0 & 1 \\ 
                               -1 & 0 
           \end{array} \right).
\end{equation}

Recently, it is pointed out by Sagan and Rubin\cite{sagan} that there exists
another solution of parameterization
\begin{equation}
A = \left( \begin{array}{clcr} I \cosh\phi & {\bar w} \sinh\phi \\ 
                               -w \sinh\phi & I \cosh\phi 
           \end{array} \right)
    \left( \begin{array}{clcr} s_1 & 0 \\ 
                               0 & s_2 
           \end{array} \right)
\end{equation}
where $\det w = -1$.

In the case of a strongly coupled lattice, for example, the interaction 
region of the Low Energy Ring(LER) of PEP-II, both solutions are needed for 
a complete parameterization of the region. 

\begin{figure}[htb]
\centering
\includegraphics*[width=60mm]{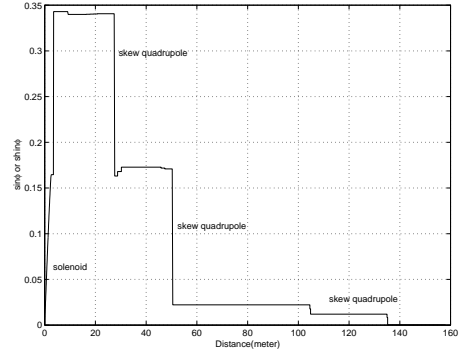}
\caption{Coupling angle in a half of the interaction region of the LER}
\label{fig:coupling}
\end{figure}

Here we choose $\phi$ as a parameter of linear coupling because it has 
two very important properties: System is decoupled when $\phi = 0$ 
and $\phi$ can be changed only by a skew quadrupole or solenoid as 
shown in Fig.~\ref{fig:coupling} from which the locations of the 
skew quadrupoles and solenoid are clearly seen. 

\section{BEAM PROFILE}

In the normalized coordinate, the sigma matrix can be written as
\begin{equation}
\epsilon = \left( \begin{array}{clcr} I \epsilon_1 &  0 \\ 
                                      0 & I \epsilon_2 
           \end{array} \right)
\end{equation}
where $\epsilon_1$ and $\epsilon_2$ are the equilibrium emittances in the
eigen planes ignoring the off-diagonal terms at the order of the
damping increment per turn. The sigma matrix in the physical coordinate
is obtained with the transformation
\begin{equation}
\Sigma = A \cdot \epsilon \cdot A^T.
\end{equation} 
The corresponding equilibrium Gaussian distribution is
\begin{equation}
\rho(x,P_x,y,P_y)={1\over2\pi {(\det\Sigma)}^{1\over2}}\exp(-{1\over2} 
                  Z^T \cdot \Sigma^{-1} \cdot Z)
\end{equation}
where $Z$ is the vector of canonical coordinates. The beam profile
in the configuration $x$ and $y$ is derived by integrating the canonical
momentum $P_x$ and $P_y$
\begin{equation}
\rho(x,y) = \int_{-\infty}^\infty \int_{-\infty}^\infty 
            dP_x dP_y \rho(x,P_X,y,P_y). 
\end{equation}

\begin{figure}[htb]
\centering
\includegraphics*[width=60mm]{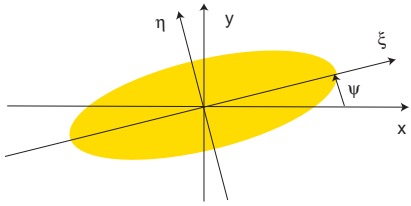}
\caption{Beam profile}
\label{fig:rot}
\end{figure}

The result of the integration is
\begin{equation}
\rho(x,y)={{(\det K)}^{1\over2}\over2\pi}
\exp(-{1\over2} z^T \cdot K \cdot z)
\end{equation}
where $z$ is the vector of configuration coordinates namely
\begin{equation}
z = \left( \begin{array}{cc} x  \\ 
                             y  
           \end{array} \right)
\end{equation}
and $K$ is the inverse of the sigma matrix
\begin{equation}
K = {1\over (\sigma_{xx}\sigma_{yy} - \sigma_{xy}\sigma_{yx}) }  
           \left( \begin{array}{clcr} \sigma_{yy} &  -\sigma_{xy} \\ 
                                      -\sigma_{yx} &  \sigma_{xx} 
           \end{array} \right).
\end{equation}
Moreover, the symmetric matrix $K$ can be diagonalized by a rotation 
transformation as shown in Fig. \ref{fig:rot} 
\begin{equation}
     \left( \begin{array}{cc} \xi  \\ 
                              \eta  
           \end{array} \right)
  =  \left( \begin{array}{clcr} \cos\psi &  \sin\psi \\ 
                               -\sin\psi &  \cos\psi 
           \end{array} \right)
     \left( \begin{array}{cc} x  \\ 
                              y  
           \end{array} \right),
\end{equation}
where $\psi$ is the tilted angle of the beam. The parameters that
describe the beam in two coordinate systems relate with each other
as following
\begin{eqnarray}
K_{xx} &=& ({\cos^2\psi\over\sigma_\xi^2} + {\sin^2\psi\over\sigma_\eta^2}) 
     \nonumber \\
K_{xy} &=& ({1\over\sigma_\xi^2} - {1\over\sigma_\eta^2})\sin\psi\cos\psi 
     \nonumber \\
K_{yy} &=& ({\sin^2\psi\over\sigma_\xi^2} + {\cos^2\psi\over\sigma_\eta^2})
\label{eqn:geom}
\end{eqnarray}
and 
\begin{eqnarray}
\tan2\psi &=& {2K_{xy}\over K_{xx} - K_{yy}}  \nonumber \\
{1\over\sigma_\xi^2} &=& {1\over2} [ (K_{xx} + K_{yy}) + {K_{xx}-K_{yy}
                     \over\cos2\psi}] 
     \nonumber \\
{1\over\sigma_\eta^2} &=& {1\over2} [ (K_{xx} + K_{yy}) - {K_{xx}-K_{yy}
                     \over\cos2\psi}]
\end{eqnarray}
where $K_{xx}, K_{xy}, K_{yy}$ are the elements of $K$ and $\sigma_\xi$
and $\sigma_\eta$ are the beam size along the long axis $\xi$ and short 
axis $\eta$ of the ellipse respectively.

\section{LUMINOSITY}

For simplicity, we ignore the effect of a finite bunch length. Given the 
two beam profiles, $\rho_1$ and $\rho_2$, the luminosity can be written as
\begin{equation}
L = n_b f_0 N_1 N_2\int_{-\infty}^\infty \int_{-\infty}^\infty
               \rho_1(x,y)\rho_2(x,y)dxdy
\end{equation}
where $n_b$ is the number of the colliding bunches, $f_0$ is the revolution
frequency, and $N_1, N_2$ are the number of charges in each position
and electron bunch respectively.

At a low beam current, the beam distribution is nearly Gaussian. Using
Gaussian as an approximation, we can evaluate the overlapping integral
\begin{eqnarray}
L_{2D} &=& \int_{-\infty}^\infty \int_{-\infty}^\infty
             \rho_1(x,y)\rho_2(x,y)dxdy \nonumber \\
       &=& {\sqrt{\det K_1} \sqrt{\det K_2} \over 2\pi\sqrt{\det(K_1+K_2)}}.
\label{eqn:dets}
\end{eqnarray} 
In order to analyze the result of the calculation, we rewrite the 
integral in terms of the geometrical parameters, $\sigma_\xi$, $\sigma_\eta$, 
and $\psi$ by substituting Eq.~\ref{eqn:geom} into Eq.~\ref{eqn:dets}
\begin{equation}
L_{2D} = L_0 {1\over \sqrt{1+e_{12}\sin^2(\psi_2-\psi_1)}}
\label{eqn:lumi}
\end{equation} 
where
\begin{equation}
L_0 = {1 \over 2\pi \sqrt{\sigma_{\xi1}^2+\sigma_{\xi2}^2}
                    \sqrt{\sigma_{\eta1}^2+\sigma_{\eta2}^2} }
\end{equation}
and
\begin{equation}
e_{12} = {(\sigma_{\xi1}^2 - \sigma_{\eta1}^2)
          (\sigma_{\xi2}^2 - \sigma_{\eta2}^2) \over
          (\sigma_{\xi1}^2 + \sigma_{\xi2}^2)
          (\sigma_{\eta1}^2 + \sigma_{\eta2}^2) }.
\end{equation}
Clearly, the luminosity is at its maximum when $\psi_2 = \psi_1$ which
is always the case in a symmetric collider due the symmetries. When two 
flat beams have identical size, we have  
\begin{equation}
e_{12} \approx {\sigma_\xi^2 \over 4\sigma_\eta^2}.
\end{equation}
It is easy to see that the reduction of the luminosity due to the 
difference of the two tilted angles are much enhanced by the aspect ratio
$\sigma_\xi$/$\sigma_\eta$ of the beams.

\section{MEASUREMENT}

The luminosity depends upon all six geometrical parameters $\sigma_{\xi1}, 
\sigma_{\xi2}, \sigma_{\eta1}, \sigma_{\eta2}, \psi_1$, and $\psi_2$ which
describe the size and orientation of the beams. It is impossible to extract
these parameters directly from the luminosity alone. By transversely scanning 
the beam crossing the other one, we could extract more constraints
among them.  

Let's calculate the overlapping integral with an offset of centroid of a beam
\begin{equation}
L_{2D}(x_0, y_0) = \int_{-\infty}^\infty \int_{-\infty}^\infty
              \rho_1(x-x_0, y-y_0)\rho_2(x,y)dxdy.
\end{equation}
After a lengthy but straightforward algebra, we obtain
\begin{equation}
L_{2D}(x_0, y_0) = L_{2D} \exp(-{1\over2} z_0^T \cdot M \cdot z_0).
\label{eqn:scan}
\end{equation}
The result agrees with Eq.~\ref{eqn:lumi} when $x_0=y_0=0$. $M$ is
a 2$\times$2 symmetric matrix that has the following elements
\begin{eqnarray}
M_{xx} = {1\over D}(a_1K_{1xx}+a_2K_{2xx}) \nonumber \\
M_{xy} = {1\over D}(a_1K_{1xy}+a_2K_{2xy}) \nonumber \\
M_{yy} = {1\over D}(a_1K_{1yy}+a_2K_{2yy})
\end{eqnarray}
where $a = \sigma_\xi^2\sigma_\eta^2$ and
\begin{equation}
D = (\sigma_{\xi1}^2+\sigma_{\xi2}^2)(\sigma_{\eta1}^2+\sigma_{\eta2}^2)
    [1 + e_{12} \sin^2(\psi_2-\psi_1)].
\end{equation}

Please note that $L_{2D}(x_0,y_0)$ is again Gaussian and is normalized 
to unity
\begin{equation}
\int_{-\infty}^\infty \int_{-\infty}^\infty L_{2D}(x_0,y_0)dx_0dy_0 = 1.
\end{equation}
Thus, similar to the treatment of the single beam profile, we can diagonalize 
$M$ with a rotation $\Psi$ \begin{eqnarray}
\tan2\Psi &=& {2M_{xy}\over M_{xx} - M_{yy}}  \nonumber \\
{1\over\Sigma_\xi^2} &=& {1\over2} [ (M_{xx} + M_{yy}) + {M_{xx}-M_{yy}
                     \over\cos2\Psi}] 
     \nonumber \\
{1\over\Sigma_\eta^2} &=& {1\over2} [ (M_{xx} + M_{yy}) - {M_{xx}-M_{yy}
                     \over\cos2\Psi}],
\end{eqnarray}
where we denote $\xi$ and $\eta$ as the principal axes.

As an experiment, we move one beam against the other one horizontally 
with a closed orbit bump at the collision point and measure the luminosity
as a function of the offset. The luminosity of the scan is proportional to
\begin{equation}
       L_{2D} \exp (-{x_0^2\over2 \Sigma_x^2})
\end{equation} 
where 
\begin{equation}
\Sigma_x = {\sqrt D\over \sqrt{a_1K_{1xx} + a_2 K_{2xx}}}
\end{equation}
To simplify the calculation, let's assume that $\psi_2 = \psi_1 = \psi$
which is a very good approximation when the luminosity is well optimized.
For flat beams and small $\psi$'s, we have 
\begin{equation}
\Sigma_x \approx \sqrt{\sigma_{\xi1}^2 + \sigma_{\xi2}^2}
           [ 1 - {\sigma_{\xi1}^2 + \sigma_{\xi2}^2\over 2
                 (\sigma_{\eta1}^2 + \sigma_{\eta2}^2)} \psi^2].
\end{equation}  
We can see that $\psi$ makes $\Sigma_x$ smaller than the design value
and the effect of $\psi$ is much enhanced by the aspect ratio. 

Similarly, we carry out the calculation for the vertical scan
\begin{equation}
\Sigma_y \approx \sqrt{\sigma_{\eta1}^2 + \sigma_{\eta2}^2}
           ( 1 + {1\over2}\psi^2),
\end{equation}
where $\psi$ makes $\Sigma_y$ slightly larger but is not enhanced by the 
aspect ratio as in the horizontal scan.

\begin{figure}[htb]
\centering
\includegraphics*[width=60mm]{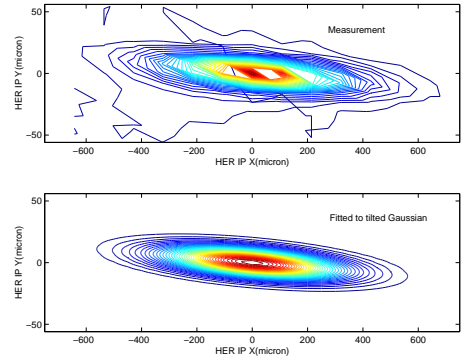}
\caption{Two dimensional luminosity scan for PEP-II}
\label{fig:scan}
\end{figure}

More general, we can scan the luminosity as a function of both 
horizontal and vertical offsets in a two-dimensional grid. 
The result of the measurement is shown in Fig.~\ref{fig:scan} 
as contour plots of the specific luminosity.

The results of the fitting are $\Sigma_\xi = 205$ $\mu$m, $\Sigma_\eta = 7.5$ 
$\mu$m, and $\Psi = -1.13^0$ compared with the design values: 
$\Sigma_\xi = 219$ $\mu$m, $\Sigma_\eta = 6.64$ $\mu$m, and $\Psi = 0$.

\section{CONCLUSIONS}
We have calculated the degradation of luminosity due to different tilted
angles of colliding beams. For same tilted angles, the higher the aspect 
ratio is the more luminosity reduction will be. Furthermore, we have 
computed a general luminosity formula for off-centered beams. The 
formula is used to understand the luminosity scan, especially for the 
two-dimensional scan. 

\section{ACKNOWLEDGEMENTS}
We would like to thank W. Kozanecki, M. Minty, J. Seeman, and U. Wienands  
for many helpful discussions.

\end{document}